# Gauge symmetry, left-right asymmetry and atom-antiatom systems: Coulomb's law as a universal molecular function.

G. Van Hooydonk, Department of Library Sciences and Department of Physical & Inorganic Chemistry, Ghent University, Rozier 9, B-9000 Ghent, Belgium (Email: guido.vanhooydonk@rug.ac.be)

We prove that Coulomb's 1/R-law is the universal function, needed for scaling in molecular spectroscopy. To obtain this result we introduce intra-atomic charge inversion. This generates an algebraic switch in the Hamiltonian of 4-particle systems when going from atom-atom XX to atom-antiatom X$\underline{X}$ systems. This switch is a consequence of atomic left-right asymmetry (handedness, chirality). For X$\underline{X}$ systems, this parity operator can reduce the 10 term Hamiltonian to 1 or 2 terms. The reduced Hamiltonian reproduces and scales potential-energy curves (PECs) of normal bonds. For 9 bonds (HH, HM and MM, M alkali metal) observed levels and turning points of XX systems coincide with those calculated for X$\underline{X}$ systems. Hydrogen-antihydrogen reactions, feasible in the near future, will produce a normal HH molecule. If true, this would solve the problem about the existence of antimatter. Nature prefers charge anti-symmetry in neutral bound 2-particle systems. In neutral bound 4-particle systems dipole anti-symmetry is preferred, which implies that charges in two interacting neutral particles are not assigned according to convention.

## 1. Introduction.

In the 2$^{nd}$ quarter of the 20$^{th}$ century there was a parallel development in physics and chemistry. With parity violation discovered around 1950 a break occurred. This led to the new physics, the Standard Model (and beyond), whereas in chemistry mainly computational procedures for many particle systems were refined [1a]. Still, physics and chemistry remain united in solving 4-particle systems, important in the advent of hydrogen-antihydrogen reactions. H$_2$ is the simplest stable 4-particle system with unit charges, the ultimate test case for elementary particle theories. Quantum mechanics even describes H$_2$ exactly [1a]. But interest remains in classical and semi-classical approximations for various reasons, not only because of the complexity of quantum mechanical procedures for large systems but also because of the study of chaos in quantum systems and the correspondence principle. In this context, Bohr's molecular bonding models [1b] were reanalyzed [1c]. In both quantum mechanical and classical methods, the 4-particle Hamiltonian is rather complex [1a,1c] and the way in which charges are distributed is critical for system-stability deriving from Coulomb forces [1c, 1d]. For instance, the computation of 2-electron interactions is the bottleneck of quantum chemistry (the Coulomb problem [1e]). To avoid computational problems, Coulomb-attenuated HFS calculations have been proposed, originating from a suggestion to cut off the long range branch of a Coulomb potential [1f]. The present work focuses on unconventional if not chaotic charge distributions in 4 particle systems, without altering Coulomb forces.

If leptons have mass $m_a$, $m_b$, nucleons $M_1$, $M_2$, the 10 term Hamiltonian is

$$H = \tfrac{1}{2}m_a v_a^2 + \tfrac{1}{2}m_b v_b^2 + \tfrac{1}{2}M_1 v_1^2 + \tfrac{1}{2}M_2 v_2^2 - e^2/R_{1a} - e^2/R_{1b} - e^2/R_{2a} - e^2/R_{2b} + e^2/R_{ab} + e^2/R_{12} \quad (1)$$

The standard premise in chemistry [1a] that leptons have negative, nucleons positive charges secures that nucleon-lepton interactions account for bonding as in Heitler-London (H-L) theory [2]. However, (1) as it stands does not give a hint about two basic issues:
(i) singlet-triplet splitting observed invariantly, and
(ii) shape/scale invariance of potential-energy curves (PECs). Triplet PECs follow a repulsive Coulomb law, singlet PECs show left-right asymmetry at the minimum but the function is unknown.

Both problems are connected but only (i) was solved satisfactorily in H-L theory [2]. Their solution derives from symmetry effects of both spin and wave function (Pauli's fermion anti-symmetry) in a wave mechanical framework. The wave equation with (1) must be solved first and atomic energies are then subtracted to get bond details (PECs). The complex H-L solution for (i) prohibits a simple one for (ii), shape/scale invariance of PECs. For singlet states, this can only be accounted for by means of a universal function, still to be found and, according to Tellinghuisen [3], this is the *Holy Grail of Molecular Spectroscopy*. PEC invariance is reflected in part [3,4] in a simple behavior of spectroscopic constants $\omega_e$, $B_e$, $\omega_e x_e$ and $\alpha_e$. Hamiltonian (1) can not lead to shape/scale invariant PECs, unless it is reduced to a function with just $R_{12}$ as a variable. For singlet states, empirical 1/R-potentials can indeed account for many PECs [4] but the $R_{12}$-term in (1) is repulsive, not attractive. In addition, the best asymptote for scaling is Coulomb's asymptote $e^2/R_e$ [4], *not* the atomic dissociation limit $D_e$. This is contrary to convention, since $D_e$ is the standard scaling factor in molecular spectroscopy (the Sutherland parameter).

To solve the wave equation for (1) assumptions must be made about the molecular wave function. H-L theory uses only the VB part $\psi_{VB}$ of the complete Hund-Mulliken MO wave function

ψ_MO = ψ_VB + ψ_ION

although the ionic function, apparently, is equally important. Here, we try to solve problem (i) and (ii) starting from (1) directly without relying on external symmetry effects. A solution derives from $\psi_{ION}$, not from $\psi_{VB}$, the H-L approximation for $\psi_{MO}$.

## 2. Theory.

*a. Algebraic switch in the 10 term Hamiltonian.*

Interaction energies referring to asymptote $e^2/R_e$ and $D_e$ are instructive. Let X consist of lepton-nucleon pair $m_a$, $M_1$, Y of $m_b$, $M_2$. Pairs are charge conjugated but charge symmetry is broken by the large particle mass difference (m/M=1/1836 for H).

With the *non-Coulomb* asymptote $D_e$, (1) leads to
$$V(R)=H_{XY}=H-(H_X+H_Y)=$$
$$-e^2/R_{1b}-e^2/R_{2a}+e^2/R_{ab}+e^2/R_{12} \qquad (2)$$
or 4 terms to cope with a pair of two *neutral atoms*.

The ionic *Coulomb* asymptote gives 3 terms
$$V'(R)=H_{X+Y-}=H-(H_{X+}+H_{Y-})=-e^2/R_{1a}-e^2/R_{1b}+e^2/R_{12} \quad (3)$$
for a pair of two *charge conjugated ions*. Here, charge symmetry is *not* broken by mass difference, which is small (order 2/1836 for H).

In both (2) and (3), the internuclear term remains repulsive. At least here a switch in sign is needed for singlet states to obtain a function starting off as an attraction at the asymptote with variable $R = R_{12}$.

Switching signs of Coulomb terms corresponds with switching fermion chiralities, a result of charge conjugation in combination with the particle hole transformation [5]. Out of the 6 Coulomb terms in (1), 4 switch sign when antiatom $\underline{X}$ ($m^+M^-$) replaces atom X ($m^-M^+$). Charge invariance secures that such a switch leaves the (intra-) atomic energy invariant. Asymptotes for all 4 states XY, X$\underline{Y}$, $\underline{X}$Y and $\underline{XY}$ are identical. This has an important consequence when a switch[1], a parity operator p is introduced in (1)
$$H_p=(\tfrac{1}{2}m_av_a^2+\tfrac{1}{2}m_bv_b^2+\tfrac{1}{2}M_1v_1^2+\tfrac{1}{2}M_2v_2^2-e^2/R_{1a}-e^2/R_{2b})$$
$$+(-1)^p(-e^2/R_{1b}-e^2/R_{2a}+e^2/R_{ab}+e^2/R_{12}) \qquad (4a).$$
Terms between the first pair of brackets reflect intra-atomic charge invariance. For the remaining 4 terms, p=0 gives the classical system XX or $\underline{XX}$, p=1 gives $\underline{X}$X or X$\underline{X}$. If asymptotes are really charge invariant, p explains splitting and solves problem (i), since
$$H_p = H_{\pm} = H_0 \pm V(R) \qquad (4b)$$
The first 6 terms in (4a) belong to charge invariant asymptote $H_0$, the remaining 4 are interactions V(R)
$$V(R)_{XX} = V(R)_{\underline{XX}} = -V(R)_{X\underline{X}}$$
$$V(R)_{\underline{X}X} = V(R)_{X\underline{X}} \qquad (4c)$$
Two pairs of degenerate states appear, one being dipole symmetric (↑↑or↓↓), the other anti-symmetric (↑↓or↓↑) just like with spins. The approach differs from H-L theory but Pauli-matrices apply to both, as spin and dipole symmetries are similar. Spin (±½) and charge (± 1) operators only differ by a factor 2.

With non-Coulomb asymptote $D_e$, *intra-atomic charge inversion* p = 1 gives, instead of (2)
$$V(R)=H_{\underline{X}Y}=H-(H_{\underline{X}}+H_Y)=$$
$$+e^2/R_{1b}+e^2/R_{2a}-e^2/R_{ab}-e^2/R_{12} \qquad (5a)$$
With the ionic asymptote, (3) transforms in
$$V'(R)=H_{\underline{X}-Y+}=H-(H_{\underline{X}-}+H_{Y+})=$$
$$-e^2/R_{1a}+e^2/R_{1b}-e^2/R_{12} \qquad (5b)$$
where $Y^+$ is a composite 3 particle antianion. Both (5a) and (5b) give attraction for the $R_{12}$-term.

A solution with an algebraic switch is generic and independent of the system's unknown geometry. We will not reconsider classical Bohr models [1b, 1c].

Intra-atomic charge inversion leads to 4 states with different dipole alignments, useful for classical analysis at long range (related spin states have only minor energetic consequences).

First, (2) and (5a) approach a charge invariant asymptote $D_e$ symmetrically, conforming to (4b). If (2) reaches $D_e$ from the repulsive, (5a) reaches it from the attractive side or vice versa. With atomic radius d=½$R_{ab(e)}$=½$R_e$, dipole-dipole interactions (the magnet metaphor[1]) give
$$V(R)_{XX} = V(R)_{\underline{XX}} = +\tfrac{1}{4}(e^2/R_e)(R_e/R)^3...$$
$$V(R)_{\underline{X}X} = V(R)_{X\underline{X}} = -\tfrac{1}{4}(e^2/R_e)(R_e/R)^3... \qquad (6a)$$
varying as $R^{-3}$. Splitting is twice as large. Result (6a) applies to long range only. With respect to $D_e$, this naive Coulomb treatment gives repulsion for dipole symmetric states and attraction for dipole anti-symmetric states. Approximating chiral effects by rotating one dipole by 180° gives the opposite result in first order[2] but this is not exactly the same as a mirror symmetry effect with charge inversion (p=1).

Second, (3) and (5b) behave similarly. Ionic models have the advantage of their simple geometry. Long *and* short-range interactions obey Coulomb's law. Equating ion attraction with (3) gives $-e^2/R_{12}=-e^2/R_{1a}-e^2/R_{1b}+e^2/R_{12}$ for p=0. At short-range, $e^2/R_{12}= e^2/R_{1a}$, if $e^2/R_{1a}=e^2/R_{1b}$. This is wrong by a factor 2. For p=1 (5b) gives $-e^2/R_{12}=-e^2/R_{1a}+e^2/R_{1b}-e^2/R_{12}$ or $e^2/R_{12} \approx e^2/R_{12}$, the correct ionic Coulomb attraction for all R-values even close to $R_e$.

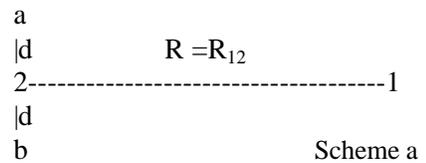

```
    a
    |d            R =R_12
    2------------------------------------1
    |d
    b                           Scheme a
```

---

[1] This switch can be visualized by 2 hand-held permanent magnets (atomic EDMs). A parallel (p=0) or anti-parallel (p=1) alignment of magnets reflects mirror symmetry and is felt when the distances between aligned magnets vary.

[2] Dipole rotation by π gives $V_p(R)=-\tfrac{1}{4}(-1)^p(e^2/R_e)(R_e/R)^3$. In $2^d$ order, mirror symmetry based states in (6a) are lower by a term in $(e^2/R_e)(R_e/R)^5$. Linear dipole alignments always give zero energy in first order.

We can refine this with ionic models (as in Scheme a) where $R_{ab}=2d$ is perpendicular to R.

Without polarization, we have for p=0 and p=1
$V'(R)_{X^+X^-} = V'(R)_{\underline{X}^+\underline{X}^-} = -(e^2/R)(1-(d/R)^2)$
$V'(R)_{\underline{X}^+X^-} = V'(R)_{X^+\underline{X}^-} = -e^2/R$ (6b).
With p=0, lepton-nucleon polarization is hampered by inter-lepton repulsion. For p=1, lepton-nucleon interactions cancels exactly. Although ionic states are always attractive by definition, a small p-dependence is observed.

For $R_{ab}$ parallel to R, we obtain respectively
$V'(R)_{X^+X^-} = V'(R)_{\underline{X}^+\underline{X}^-} = -e^2/R$
$V'(R)_{\underline{X}^+X^-} = V'(R)_{X^+\underline{X}^-} = -(e^2/R)(1+2d/R)$ (6c)

In ionic cases (6b, 6c), p-effects are subtle. Yet, a charge inverted ionic state is always low. These are easily verifiable consequences of a parity operator in Hamiltonian (4a). The covalent case is ambiguous but only in first order[2] (see also below).

Charge inversion solves problem (i) generically but for problem (ii) additional information is needed. From both long and short-range interactions in ionic models (6b, 6c), the solution about stability is unambiguous. Unfortunately, it is impossible to get an extremum. The most critical issue is a minimum for the lowest charge inverted ionic $\underline{X}^+X^-$ or $X^+\underline{X}^-$ state in (6c). Whether or not a minimum exists is uncertain. BOA (Born-Oppenheimer Approximation) may apply but this is not evident. Comparing V'(R), dV(R)/dR is largest for (6c). Unlike charge inverted states $V(R)_{\underline{X}X}$[3], the p=1 state in (6c) will intersect asymptote $D_e$ at large R and reach the extremum, if it exists, before any other state. Extrapolating long-range attraction to short range, gives the p=1 case in (6c) as most stable state. But for the simple case HH, the character of $H\underline{H}$ is still an open question. Since data on H---$\underline{H}$ reactions could be available soon a solution is needed. But even sophisticated quantum mechanical methods are controversial about $H\underline{H}$.

*b. Controversy about $H\underline{H}$.*

Modern quantumchemical methods for 4-particle systems have *ab initio* status [1a] and reach spectroscopic accuracy. Then it is surprising there is a controversy about $H\underline{H}$. The Richard group [6-8] suggests $H\underline{H}$ is unbound, which is confirmed by Monte Carlo simulations [9]. Abdel-Raouf and Ladik [10] claim $H\underline{H}$ is bound. Computational efforts are great and both methods seem reliable. The origin of the controversy lies in the effect p in (4a). Perhaps, something is wrong with the premises of quantum chemistry (charges, wave functions, correlation[4]), despite its successes [1a].

We showed above that p=0 and p=1 solutions for (4a) reach the same asymptote $D_e$ from different sides. If HH is bound, $H\underline{H}$ is unbound as in the first thesis [6-9]. A different spatial lepton configuration can invert this result but only in first order[2]: maybe this minor difference is a classical explanation for the origin of wave mechanical ambiguity with $H\underline{H}$. Even at long range, a charge inverted atomic state is always below a dipole rotated normal state. In the cold atom region, $R_e$/R-values equal to 1/20 and 1/10 respectively give small energy differences of only <0.01 and <0.4 cm$^{-1}$ if C = 110,000cm$^{-1}$ in favor of charge inverted p=1 states. Also all ionic states lead to enhanced attraction for $H\underline{H}$, supporting Abdel-Raouf and Ladik's thesis [10].

Conventional HH is less stable than its charge inverted version $H\underline{H}$, with H and $\underline{H}$ related by mirror or left-right asymmetry. This is not a spin and orbital symmetry but a dipole symmetry effect, completely absent as such in conventional H-L theory.

These are all classical easily verifiable results deriving from long range behavior and dipole-dipole Coulomb interactions. The extrapolation to short range is critical. As for problem (ii), the extremum, if any, for $X\underline{X}$ and in particular for $H\underline{H}$ must be found.

*c. Reducing the 10 term 4-particle p=1 Hamiltonian to a Kratzer and a Coulomb potential.*

For p=1, lepton-nucleon interactions in (4a) can cancel exactly for an unknown geometry. In this *ad hoc* hypothesis and *independent of asymptote choice*, any p=1 solution for (4a) invariantly gives
$H_{p=1}=(\frac{1}{2}m_av_a^2+\frac{1}{2}m_bv_b^2-e^2/R_{ab})$
$\quad\quad +(\frac{1}{2}M_1v_1^2+\frac{1}{2}M_2v_2^2-e^2/R_{12})$ (7).
This is either positronium-protonium or hydrogen-antihydrogen, depending on the value of the reduced mass (central force character, BOA or geometry of subsystems). In the limit, a mass-less 4-unit charge system results. Without the term in $R_{12}$, (7) applies to anion/helium like 3-particle systems. With electronic wave functions, the $R_{12}$-term may be disregarded first and added at the end of the procedure.

Instead of sophisticated quantum mechanical methods which lead to a controversy for $H\underline{H}$ [6-10], we rely on easy to check solutions for (7) to solve problem (ii) and to finalize the stability problem about $X\underline{X}$. Two solutions exist, which must be confronted with observation.

a. First, assume (7) is a generic result and that leptons can be redistributed over nucleons in such a way that two unspecified quasi-central force systems emerge (BOA). In practice, this configuration is restricted to $R_e$-domain in VB schemes. If R ≈ $R_e$, then R ≈ $R_{ab}$ also and a first solution for (7) can be
$H_{p=1} \approx 2(\frac{1}{2}\mu v^2 - e^2/R)$ (8)

---
[3] This state will deviate from $D_e$ at much slower rate (6a).
[4] With a James-Coolidge procedure for correlation $1/R_{ab}$. Some of these important computational difficulties with electron correlation are discussed in Ref. [1a, 1e, 1f].

a Bohr-like model [1b] *with charge inversion*. Here, μ is a reduced mass and the asymptote is situated between zero and the total well depth. Eqn. (8) is a reduction from 10 to 2 terms in Hamiltonian (1). In contrast with long-range forces above, (8) is valid at short range, the $R_e$-domain, where the 4 lepton-nucleon interactions can cancel exactly[5]. Then, one can say something about the *dynamics* of the system only by assuming the lepton's angular momentum is constant. Bohr's quantum condition secures that

$$\mu v^2 \sim B/R^2 \qquad (9)$$

where B is a constant. Its value depends on $R_e$. With asymptote A, a Kratzer potential [11] is the result

$$H_{p=1} = -A + 2(B/R^2 - e^2/R) \qquad (10)$$

This potential has a classical minimum when the derivative with respect to R vanishes or when

$$B = \tfrac{1}{2}e^2 R_e \qquad (11).$$

Result (11) leads to the eigenvalue

$$H_{p=1}(R_e) = -A - e^2/R_e \qquad (12)$$

a Coulomb asymptote, anchored at A between zero and the total well depth. Kratzer's potential (10) was proposed already in 1920, after Bohr's atom theory but before Schrödinger's wave mechanics. It is itself a generalized Bohr *atom* formula but it is also very useful for *molecules* [4, 12] in another generalized form due to Varshni [13]. Kratzer-Varshni's potential scales molecular *spectroscopic constants* efficiently [4]. It accounts for PECs only up to $D_e$[14], which is crosses at the critical distance since $e^2/R_e \gg D_e$.

The usefulness of (10) shows in oscillator form

$$W(R) = (e^2/R_e)(1 - R_e/R)^2 = Ck^2 \qquad (13)$$

Here k is the Kratzer variable[6]

$$k = 1 - R_e/R \qquad (14)$$

and C is the *ionic* Coulomb asymptote. Form (13) appears naturally for (7), though no assumption was made about asymptotes. In fact, equations (9) and (10) are confined to the $R_e$-domain (see above).

This first solution for X$\underline{X}$ systems may require *ad hoc* hypotheses about an unknown geometry for the system but an acceptable oscillator model and asymptote are obtained. This result is consistent with spectroscopic data [4, 14-18] for diatomic bonds XX, i.e. for *atom-atom* systems in H-L sense. However, solution (13) can only apply for a bound singlet state (p=1) and there is no information about the triplet state. Nevertheless, it throws another light on the controversy about the stability of H$\underline{H}$: it seems this is not only the more stable version of HH but it should also have a minimum.

If true, H-L theory is only a complex way for introducing an atom-antiatom switch in (1) as in (4a) to describe what we now call a chemical bond. Confronting (13) with observed PECs is done below but preliminary studies show this conclusion may be unavoidable [4, 14-18].

b. A second solution derives from a 3-particle anion/helium like subsystem. The internuclear term is treated separately as in ionic central force systems, see (6b). Without polarization, the 4 lepton-nucleon interactions now *always* cancel *for all R*, not only in the $R_e$-domain as with solution a. This is the greatest advantage of ionic over covalent models but also the reason why ionic models are described as 'naive'. Ionic bonding is 200-years old (Davy and Berzelius [4]). Their naive[7] scheme uses just one Coulomb *attraction* to describe a complex 4-particle system, which seems like an oversimplification indeed.

With $\underline{X}^-$ and $Y^+$ defined above, (7) leads to

$$H_{p=1} = -A' - e^2/R = -IE_Y - EA_Y - e^2/R \qquad (15).$$

$IE_Y$ and $EA_Y$ are respectively, ionization energy and electron affinity of Y, both atomic *not molecular* constants, not varying with R.

$$A' = -IE_Y - EA_Y \qquad (16a).$$

is the eigenvalue of the 3-particle ion/He system.

By definition, a naive picture with one Coulomb interaction (15) gives the same Coulomb asymptote as in (12) at a more specified intercept A' in the total well depth. Unlike (10), (15) can *never* lead to a minimum nor a description of a repulsive triplet state. Whereas (10) gave a minimum and led to an acceptable oscillator form (13), (15) does neither, since no dynamics is involved. Taking derivatives of (15) with respect to R does not lead to an extremum. A minimum, if any, can only be generic and, if so, it must be hidden in Coulomb's law itself (see below) as must be the information about splitting towards a triplet state.

If (7) were really positronium-protonium, the energy would be

$$H_{p=1} = -\tfrac{1}{2}IE_H - e^2/R$$

with an asymptote of about 54800 cm-1. This is of the correct order of magnitude but only applicable to H. Also the problem with the minimum remains.

In (6b), the small perturbation term

$$-e^2/R_{1a} + e^2/R_{1b} \;(\approx 0) \qquad (16b)$$

was neglected. This is important to find an extremum and oscillator behavior for (15). If so, problem (ii) is also solved with the naive ionic approximation.

*d. Gauge symmetry and the generic minimum.*

Coulomb's law for 2 charges (equal masses) is

$$V_{\pm}(R) = \pm e^2/R = (-1)^t e^2/R \qquad (17)$$

where t is a parity operator, representing an algebraic symmetry for attraction and repulsion. With charge invariance, (17) led to (4c) and a degeneracy of states with symmetries governed by atomic dipoles. Coulomb's law (17) is not complete, since potentials, acting upon a unit charge, are not gauge invariant. When a 2-particle system is bound, $V_{\pm}(R_e) = \pm e^2/R_e = \pm C$. The sign of C determines the position on the axis, where (17) will be anchored. This choice of the sign for C is arbitrary and fixed by convention only.

---

[5] Configurations as in ionic schemes or in a classical Watt regulator are possible. The latter appear in Bohr's first molecular models [1b] (see also [1c]).

[6] Dunham's widely used potential based upon variable k/(1-k) can not converge. Function (13) will always reach the Coulomb asymptote C [4]. Expansions of variable k near $R_e$ lead to continued fractions also important for fractal/chaotic behavior [1d, 18].

[7] In fact, establishment immediately rejected these naive ionic models after H-L theory was available.

C-symmetry is represented with parity operator g
$$C_{\pm} = \pm C = (-1)^g C \qquad (18)$$
With g in (18) different from t in (17), degeneracy like in (4c) for dipole interactions is removed. Two-dimensional scaling of axes y (C) and x (R) by means of scaling factors $R_e$ for R and $|e^2/R_e|$ for $|C|$ is possible. Instead of (17), 4 non-degenerate states are generated generically (i.e. without convention)
$$W(R) = C_{\pm} + V_{\pm}(R) = (-1)^g C (1+(-1)^{t-g} R_e/R)$$
$$w(m) = W(R)/C = (-1)^g (1+(-1)^{t-g} R_e/R) \qquad (19)$$

The two pairs have symmetries t-g = 0 and t-g = 1, whereby gauge and interactions have the same or opposite signs. In any pair, one of the two states is charge symmetric (++ or --), the other charge anti-symmetric (+- or -+). The behavior of the states in (19) at the tree level is shown in Fig. 1[8,9].

Two states belong to a positive (C > 0), two to a negative world (C < 0). The worlds are isospectral, as algebraic symmetry applies. *Only one world can be allowed by convention* but a t-g=1 state starting in one world will extend into the other. Unlike (17), the 2 non-degenerate *attractive states* (+- and --) with t-g =1 always cross by definition, t-g=0 states never cross. Symmetry must be broken at crossing point $R_e/R=1$. With a total gap of 2C, a generic minimum $C=e^2/R_e$ is obtained. *Gauge symmetry applied to Coulomb schemes produces a generic minimum, independent of convention and dynamics*. The virial is always obeyed. Fig. 1 applies only to fermions and a relation with bonding between 2 neutral atoms (bosons) is not evident.

The two *attractive* t-g=1 states with different signs for asymptote and interaction, are
$$C - e^2/R \qquad (20a)$$
$$-C + e^2/R \qquad (20b)$$
and originate in the positive (20a) or negative world (20b) (see Fig. 1), although conventionally (20b) would be *repulsive* (--). Fig. 1 results solely from algebraic $R^n$-laws and conventions about gauges and charges. PECs deriving from n=-1 have all the characteristics of observed singlet XX PECs, except for curvatures [18].

But states with the same t - g = 1 symmetry can not cross and perturbation is required at $R_e$. Sum and difference of (20a) and (20b) are 0 and $2(C - e^2/R)$. With constant perturbation P, the perturbed Coulomb state in the positive world is
$$W(R) = C ((k^2 + p^2)^{1/2} - p) \qquad (21)$$
Here p=P/C is a reduced perturbation (not the parity operator in (4a)). This can be subtracted to obtain an oscillator presentation with zero energy at $R_e$.

At the tree level, the effect of a small constant perturbation $p^2=0,1$ (p≈0.33, see Eqn. (24) below) is shown in Fig. 2. For comparison, a Kratzer potential (13) is added but this is shifted upward with an amount p to make minima coincide. Although for (13) and (21) k-dependencies may be very different, the PECs are not. Both are consistent with shape and scale invariance of observed *atom-atom* singlet PECs [18]. Both exhibit the correct left-right asymmetry at the minimum. Nevertheless, both PECs in Fig. 2 are derived from *atom-antiatom* Hamiltonian (4a) with p=1. Both closed form analytical X<u>X</u> potentials refer to Coulomb asymptote $e^2/R_e$, so important for scaling XX PECs and constants [4, 18].

But (intra-atomic) charge invariance secured that charge inversion leaves atomic energies invariant when going from X to <u>X</u>. Fig. 1 shows that this is not so: gauge symmetry overrules charge invariance. By allowing for a (virtual) negative world, degeneracy (4c) is removed in (19), as illustrated in Fig. 1. This gives the generic minimum in (21) for Coulomb's law in the positive world (Fig. 2). But the net result is that a charge invariant asymptote in the positive world is restored, since the origin of the repulsive left branch of a PEC is to be found in attraction in the (virtual) negative world.

Gauge symmetry leads to an extremum for 2-particle systems. For 4-particle systems, the problem is more complicated, see (1) instead of (17) and (19). The perfect symmetry in Fig. 1 and 2 applies to 2 unit-charges for which it is difficult to imagine that the perturbation needed in (21) is at work, if self-perturbation is excluded. Examples are positronium and protonium. Coulomb attraction is used in full to describe the system, so it may not be used once more as a perturbation. Describing mass symmetrical 2-particle systems with gauge symmetry seems useless (annihilation, see above). Positronium is stable[10] with energy $-\frac{1}{2}IE_H$ on account of dynamics and quantum behavior.

But in a more complex neutral N-particle system the perturbation needed in (21) can be present. The neutrality condition leads to N=4 (chemical bonds),

---

[8] Extending Fig. 1 results to a second much larger gap $C_1 \gg C$, crossing at smaller $R'_e=e^2/C_1$ will result in a fine structure at $R'_e$. The opposite case also applies.

[9] The positive-negative world distinction is only formal and is easily removed by adding a large constant gauge $+C_0 \gg C$ such as the absolute mass equivalent $(m+M) c^2$.

[10] Positronium (and protonium) states have short lifetimes.

where the single Coulomb interaction for a 2-particle system is replaced by 6 as in (1). But for ionic p=1 systems a single Coulomb attraction (15) emerges which would allow us to apply a gauge symmetry based 2-particle scheme like (19) to chemical 4-particle systems. The only condition is that these are partitioned in mass asymmetrical ionic subsystems (a composite 3-particle ion and its charge conjugated non-composite antiion). In this way, the perturbation needed to break symmetry in (21) but lacking in mass symmetrical 2-particle systems, appears naturally for N=4 as in (16b). All this applies to so-called naive ionic bonding models.

Theoretically, intra-atomic charge inversion can be important for chemical bonding. The removal of the degeneracy in 2-particle systems (19) applies to 4-particle systems. In Fig. 1, the arrows represent the dipole-dipole interactions (similar to a spin notation) for 4 the states. All what has been said above about (virtual) PECs for 2-particle systems can become a reality for 4-particle systems and their PECs[11]. Anion-cation systems obey similar schemes [17,18] but the difference between normal and charge inverted anion-cation pairs are subtle (see above).

Solution b. starts from long range, see (6b, 6c), (15) and is rather vague about the minimum, where 'a' perturbation should occur. Solution a. is specific about the minimum but, nevertheless, refers to the same asymptote as solution b. Therefore it is quite tempting to confront both solutions.

Whether or not (15) is an alternative for bonding now depends on the analytical form of function P(R).

*e. Perturbed Coulomb potential.*

Classically, dipole-dipole interactions use atom polarizabilities and functions of general form $R^{-n}$ with n>>1 and/or sums of similar terms with higher n, depending on the degree of approximation (see above and Ref. [18]). Instead of applying this classical analytical solution, we shortcut the circuit and confront both solutions, as explained above. This gives the possibility to amend Kratzer's result (13) to find out whether it can be improved or not. Both give the same asymptote $e^2/R_e$.

Equating (13) with (21) will provide us with information about P(R). If $C_K$ is any Kratzer and $C_C$ any Coulomb asymptote, we obtain

$$C_K k^2 = C_C ((k^2 + p^2)^{1/2} - p) \qquad (22a)$$

With $C_K = C_C$, this leads to P(R) or p(k)

$$p = \frac{1}{2}(1-k^2) \qquad (22b).$$

This is trivial and points to the corresponding virial, as it should. Using this dependence in equation (22a) again gives

$$C_K k^2 = C_C ((k^2 + p^2(1-k^2)^2)^{1/2} - p) \qquad (23)$$

Instead of Coulomb PEC (21), the result is now

$$W(R)/C = (k^2 + (0{,}366025(1-k^2))^2)^{1/2} - 0{,}366025 \qquad (24)$$

If $C = e^2/R_e$, PEC (24) solves the problems with the one term solution b. (15): minimum and an oscillator form are obtained. PEC (24) is similar to but always *below* the Kratzer PEC (13) if both start from the same asymptote (see Fig. 2 where p = 0,33 was used, close to 0,366). Numerically, p = 0,366025 = $(3^{1/2}-1)/2$ (a more detailed derivation is in Ref. [18]). PEC (24) is an hybrid function of Kratzer solution (13) and gauge symmetry based Coulomb function (21).

PECs reaching the atomic dissociation limit $D_e$ instead of $C = e^2/R_e$ are available with

$$PEC = \frac{1}{2}(W(R)+D_e) - \frac{1}{2}((W(R)-D_e)^2 + V'(R))^{1/2} \qquad (25)$$

where W(R) can either be (13) or (24). Perturbation $V'(R)$[12] in (25) is only needed to avoid crossing at the critical distance, since $C \gg D_e$ (see above and [4]).

Both Coulomb-based results a-b, varying with k or $k^2$ for *atom-antiatom* systems, point to empirical 1/R-potentials, so valuable for interpreting PECs for *atom-atom* systems, chemical bonds in H-L theory. Both can solve problem (ii) in the Introduction. The H-L scheme can not account for the well behaving[13] empirical 1/R potentials [18], which is exactly why the search for universal 1/R-functions has been going strong for many decades [3,4,18]. If H-L theory were really complete, scaling PECs analytically would not have to be a problem [3,18] but, in practice, the scaling issue (ii) remains unsolved. This proves once more that H-L theory is not complete or at least too complicated. Both solutions a-b for (7) show that H-L theory is a cumbersome way to introduce a parity operator in (1) as in (4a). Unfortunately [1c,1d, 18], all connections with classical/semi-classical results like (13) and (15) are lost in H-L theory.

Strictly spoken, gauge symmetry would even get a solution for ionic models *without* charge inversion. But the connection with (7), generic singlet-triplet splitting and solution a. would not have been found. These are essential to arrive finally at (24) and we showed above what exactly the differences are between classical and charge inverted ionic models.

Generic gauge symmetry smoothly transforms fermion behavior (splitting at large R as in Fig. 1) into boson behavior (atoms in equilibrium around $R_e$ as in Fig. 2) without the super-potentials of SUSY. If PECs (24) are better than (13), left-right asymmetry must be the rule in nature.

---

[11] If the present scheme is validated by experiment (see below), nature would assign charges to interacting neutral particles in a way that, with respect to convention [1a], may seem chaotic in simple quantum systems (see Fig. 8 in Ref [1c] and see also [1d]). In this work, 'charge-chaos' is restricted to just a choice between atom X and charge-inverted antiatom $\underline{X}$, a mirror effect.

[12] We will use $V'(R) = 0$ throughout.

[13] Traditionally, a complete theory explains exactly why empirical relations can behave well. In this respect, H-L theory fails; this raises questions about its completeness.

*f. Ionic systems and wave functions.*

Ionic wave functions conform to the framework MO. Heitler-London-Pauling-Slater VB theory uses wave functions of general form
$\psi_{VB} = u_{1a}u_{2b} + u_{1b}u_{2a}$
LCAOs where AO is atomic wave function u.
Hund-Mulliken-Roothaan MO theory uses
$\psi_{MO} = (u_{1a} + u_{2a})(u_{1b} + u_{2b})$
$= (u_{1a}u_{2b} + u_{1b}u_{2a}) + (u_{1a}u_{1b} + u_{2a}u_{2b})$
$= \psi_{VB} + \psi_{ION}$
an equal mixture of VB and ionic structures: in MO theory $\psi_{ION}$ is as important as $\psi_{VB}$.

Ionic functions stand for the ionic configurations above. $\psi_{VB}$ and $\psi_{MO}$ are valuable molecular wave functions [1a]. Solutions with $\psi_{ION}$ instead of $\psi_{VB}$ (H-L theory) can therefore not be neglected. The difference with VB and MO theory is that for $\psi_{ION}$ gauge symmetry and atom-antiatom switches have to be introduced. These were effects not yet considered in bonding (see [17] for an early suggestion).

A generic theoretical extension of ionic Coulomb theory for diatomic molecules is that the principle of *charge alternation* will be important in the case of polyatomic molecules, which is as observed [17,18].

Finally, partitioning a 4-particle bond into a pair of charge conjugated but mass asymmetrical ions avoids pair annihilation (mass ratio 1836/1838 for H). Mass-differences of 2/1836 are so small that charge symmetry in ion-antiion pairs is not broken.

*g. Zero molecular parameter potentials.*

Using the 3 molecular parameters $R_e$, $D_e$ and $k_e$ to test the predictions with *atom-antiatom* X$\underline{X}$ PECs (13) and (24) can even be avoided. Instead of using $e^2/R_e$ ($R_e$ is a molecular constant), we extract this asymptote just from atomic data by choosing half the well depth of a bond as a substitute. This is equal to $IE_X + IE_Y + D_e \approx IE_X + IE_Y$ since $D_e/(IE_X + IE_Y) << 1$. Now, the *atomic* ionization energy $IE_X$ acts as a *molecular* asymptote for bonds XX. For bonds XY, ½ ($IE_X + IE_Y$) is a first approximation, conform to standard practices for determining atomic radii $r_X$. This leads to ionic Coulomb asymptotes since $C = e^2/R_e = e^2/2r_X = IE_X$. Errors generated by this procedure will not be too large (order 10 %).

With asymptotes deriving from atomic data only, both (13) and (24) become *zero molecular parameter functions,* whereas it is generally accepted [4,13,19] that universal molecular functions must *at least* use the 3 molecular parameters above. Just for (25), $D_e$ would be needed, but only away from the minimum.

Calculating molecular PECs from atomic data is a real challenge. The approach has *ab initio* status, as soon as it applies to 3 simple bonds $H_2$, $Li_2$ and LiH.

**3. Results and Discussion.**

PECs (RKR/IPA) for 9 bonds are analyzed: $H_2$ [20], LiH[21], KH[22], $Li_2$[23,24], KLi[25], NaCs [26], $Rb_2$[27], RbCs[28] and $Cs_2$[29]. Valence-state ionization energies are taken from NIST tables.

(a) $H_2$ (Fig. 3). The agreement is best for the generic Coulomb equation (24), although the Kratzer result (13) is still acceptable. Deviations are found at the repulsive side but the trend is as observed. For the attractive branch, the agreement is good even at long range (the important *cold atom* region [18] near $D_e$). This asymptote intersects generic curve (13) and (24). For this region, we used (25), which consumes one molecular parameter $D_e$, but only away from the minimum (about 20 % of a total PEC [18]). This first principle's molecular PEC for $H_2$ derives from H$\underline{H}$ bonding using atomic data only. It is closer to the observed $H_2$ PEC than that calculated in H-L theory, the origin of quantum chemistry. In the context of particle/antiparticle theory, Dirac's frequently cited 1929-remark, quoted recently by Pople [1a], about solving chemical problems by quantum mechanics is of interest. Dirac, the inventor of the algebraic switch we used in (4a), assumed at that time that H-L theory for 4 particles (fermions) was correct and complete. Despite the controversy in quantum mechanics about the bound or unbound character of H$\underline{H}$, we can not but say that the reaction between H and $\underline{H}$ will give a normal $H_2$ molecule[14] in the H-L sense.

(b) $Li_2$ (Fig. 4). Results for the $Cs_2$-PEC reaching $D_e$ are included. For both, agreement is better than for $H_2$ at both branches. For $Li_2$ absolute deviation for 30 turning points is 2,5% for Kratzer and only 0,6% for perturbed Coulomb potential [18], showing that PEC (24) is again better than Kratzer's (13).

(c) $Li_2$, LiH, KH, KLi, NaCs, $Rb_2$, RbCs and $Cs_2$ (Fig. 5). Here, observed level energies are plotted against theoretical ones obtained with zero molecular parameter functions. Differences are shown. Only the KH repulsive branch deviates. The slope is close to 1 and goodness of fit is high. Average deviation for 310 turning points is 9.42 %: $Cs_2$ 11.3, RbCs 10.8, $Rb_2$ 9.0, NaCs 12.2, KLi 9.3, KH 7.5 and LiH 6.3. These errors *include* long-range situations where applicable but calculated[15] with V'(R)=0 in (25). As above Coulomb PECs are better than Kratzer's.

---

[14] Although H and $\underline{H}$ are isospectral in first order, small differences can show in second order. If $H_2$ consists of H$\underline{H}$ and $\underline{H}$H, this information can be hidden in atomic spectra. Well-known fine structures in molecular spectra deriving from lepton and nucleon *spin* need not be discussed here.

[15] Part of the errors is due the use of atomic ionization energy as molecular asymptote and to the neglect of (6a).

Fig. 5 also illustrates the solution for problem (ii). We can easily scale PECs. Scale/shape invariance of 8 observed XX PECs is quantitatively accounted for with an X$\underline{X}$-scheme [18], an unprecedented result.

(d) LiH (Fig. 5). This simple bond is intermediate between $H_2$ and $Li_2$. It has the largest $D_e$-value of the 8 bonds. Both slope and goodness of fit in Fig. 5 indicate that observed data for LiH perfectly fit in an Li$\underline{H}$/$\underline{Li}$H bond scheme. When the two variables $k^2$ in (13) and generic $k_{gen}$, deriving from (24) are plotted versus level energies [18], a linear fit for $k_{gen}$ again gives the best results. Generic ionic asymptote $e^2/R_e$ is reproduced within 0,086 % of experiment. 40 calculated turning points are within 0,54 % of experiment, an average deviation of 0,012 Å. Larger deviations are found at the attractive branch. For the left branch deviation is only 0,0047 Å, very close to spectroscopic accuracy [18]. In general, left branches are easier to reproduce [3,18,30,31] than right branches where long-range forces interfere.

For KLi, deviation is even less than 0,001 Å [18]. In review [18] other examples are given. Atom-antiatom theory applied to 500 level energies and turning points for 13 bonds has a CL of 98-99 [18].

Atom-antiatom bonding theory is an overlooked atom-atom bonding theory. This is illustrated by 3 test cases $H_2$, LiH and $Li_2$ or, should we say $H\underline{H}$, Li$\underline{Li}$ and $\underline{Li}$H. Permutational symmetry requires that $X_2$ is not X$\underline{X}$ but consists of a hybrid of states X$\underline{X}$ and $\underline{X}$X. The simplicity of all analytical results and the classical concept (the magnet metaphor) contrast with complex H-L theory. PECs obey closed form formulae, in agreement with empirical evidence, an unprecedented result also.

H-L theory eventually prevented finding earlier that bonding is due to symmetry effects residing in charges within atoms. If an exchange mechanism can explain bonding classically, charge inversion in *one* atom is a realistic choice. Mirror symmetry, left-right asymmetry or chiral dipole behavior lead to antisymmetry at large. This is proven by the fact that perturbed Coulomb function (24), not obeying the virial exactly, is even better than Kratzer's.

An algebraic switch in (4a) may remove the bottleneck in quantum chemistry (see Introduction). Instead of attenuating Coulomb forces, switching their signs is a better option. In fact, gauge symmetry provides with a first principles recipe to split a single Coulomb law in two parts: a short and a long range one, the transition point being situated at $R_e$. It is not necessary to invent a computational 'trick' to reduce computer time [1e, 1f]. Disregarding either the short or long range branch of a Coulomb law is perfectly admissible in a gauge symmetry based approach. The generic PECs have an attractive Coulomb long range part $-e^2/R$ that reaches to the minimum. Here, it is replaced by a repulsive Coulomb part $+e^2/R$.

Whether or not the unconventional distribution of charges between 4 particles is a form of 'quantum chaos'[11] is not clear. In any case, our approach nicely fits in the persistent revival of classical and semi-classical approximations for small systems described well by quantum mechanics (because of chaos, the correspondence principle or both [1c, 1d]).

It is remarkable that even Coulomb's original law with strictly *unit charges* accounts well for PECs of various bonds. It is not even necessary to allow for chemical effects on charges (an otherwise legitimate parameterization procedure). This makes Coulomb's law the really universal parameter-free molecular potential, so badly needed for scaling in molecular spectroscopy [3,4,18].

### 4. Further evidence.

Atom-antiatom schemes are also consistent with a number of other observations.
1. Exceptions to orbital-symmetry based Woodward-Hoffman rules [32] in organic reactivity: conrotatory ring-closure of cis-butadiene is an example [17],
2. Dunitz-Seiler observations [33] about the *absence* of valence electron-density in the bonding region between nuclei, until today the only *experimental* indication that H-L schemes need revision [17, 33],
3. Occurrence of 5-fold symmetry in alloys [34] and the role of Euclid's golden number [35], indicating chaotic/fractal behavior in small quantum systems,
4. Absence of isotope effects in high-$T_C$ superconductors [36] and charge inversion in CuO [37],
5. Meaning and role of Cooper-pairs in general and aromatic systems in particular [38]: attraction makes lepton appearances as charge conjugated pairs their natural morphology; it is no longer difficult to explain why Cooper pairs can exist; and finally
6. The hole-concept (see above).

As outlined before [17], the interaction matrix for neutral particles X and $\underline{X}$ is

```
       | X        X
-------|------------------
   X   | XX       XX
   X   | XX       XX
```

This matrix is easily extended towards aggregates of neutral atoms X. If M is a neutral diatomic molecule $X_2$, M may replace X in this matrix. If C is a (linear) aggregate of atoms $X_n$, a neutral chain, C may replace X in this matrix and so on.

A consequence of Coulomb's law and this matrix for many particle systems is that *charge alternation* is important. We refer to, but not only to, arrays of hydrogen bridges like OH...X responsible for the coupling or breaking of two linear DNA-chains [17].

# 5. Conclusion.

Unitary symmetry, one algebraic switch in the 10 term Hamiltonian generically accounts for splitting and for scale/shape invariant PECs. *Atom-antiatom* bonding explains classically and quantitatively *atom-atom* bonding. Theory is in agreement with universal scaling [4,18], a problem in molecular spectroscopy [3,4,18,19]. The classical intuitive 200-year-old ideas of Davy and Berzelius about (ionic) bonding are still valid today if gauge symmetry and charge inversion is considered. Coulomb's law and gauge symmetry justify the 'trick' invented to remove the bottleneck in quantum chemistry [1e, 1f].

The capacity of a Coulomb asymptote to scale spectroscopic constants is impressive [4] and shows that X$\underline{X}$ bonding schemes have universal character, deriving from first principles. Coulomb's law with unit charges is a truly universal molecular function.

Gauge symmetry for $1/R$-laws provides with a simple recipe for a transition from fermion (splitting) to boson (oscillator) behavior.

Hydrogen-antihydrogen reactions, feasible in the near future [39], will probably produce normal $H_2$. This reaction is crucial for 20$^{th}$ century physics on discrete symmetries. X$\underline{X}$ systems are bound and stable. Wave mechanics is still controversial on H$\underline{H}$ stability, which remains difficult to understand if the $H_2$-PEC can be calculated with great accuracy [40]. No-extremum theories for H$\underline{H}$ [41] are probably incorrect [35] because of gauge symmetry.

Atomic symmetries like handedness, chirality or left-right asymmetry must be reconsidered at atomic and at molecular energy levels to look for higher order effects of gauge symmetry and charge inversion, not discussed[8,14]. Many experiments are under way in atomic physics because of implications for the Standard Model (and beyond) and in molecular physics the study of cold atoms is important (references are given in [18]).

But even cosmology, baryogenesis [42] and super-unification are at stake if antimatter is really present in nature [17,18,35]. The perfect (algebraic) equilibrium in the X$\underline{X}$ pair production process in chemistry gives exactly equal amounts of matter and antimatter, if atoms stand for matter and antiatoms for antimatter. We have now found that molecular singlet and triplet PECs are witnesses for the presence of antiatoms in bonds (matter). A dipole-symmetry conjugated charge-neutral and mass-positive particle pair (atom/antiatom) does not (have to) annihilate, although its 6 Coulomb terms refer to pairs of charged particles and antiparticles fermions in Dirac sense. Since a Coulomb function is even better than Kratzer's, this conclusion is unavoidable.

The gap $e^2/R_e$ covered by chemical interactions is $½\alpha^2 mc^2$, where $\alpha$ is the fine structure constant. This is $½(1/18,800)(1/1,836)$ or 0.2 ppm of the gap $m_H c^2$ covered by H-annihilation. Chemistry may be just a fine structure effect[8,9] but the underlying mechanism is universal and applicable to larger or smaller gaps if the scheme is scale invariant [18]. Extensions to other fields are given elsewhere [18].

Like all living material, man is also made up from composite neutral particle pairs. Algebraic and super-symmetry are not only important for molecules but also for cell biology, especially for electrostatic interactions between pairs of DNA-chains [17, 43].

**Acknowledgments.**

We are indebted to R. L. Hall, J. R. Le Roy, W. C. Stwalley, J. Tomasi, L. Wolniewicz and especially to Y. P. Varhsni (data, pre-/reprints, correspondence).

---

Fig. 1 (Fig. 5 from ref [18])

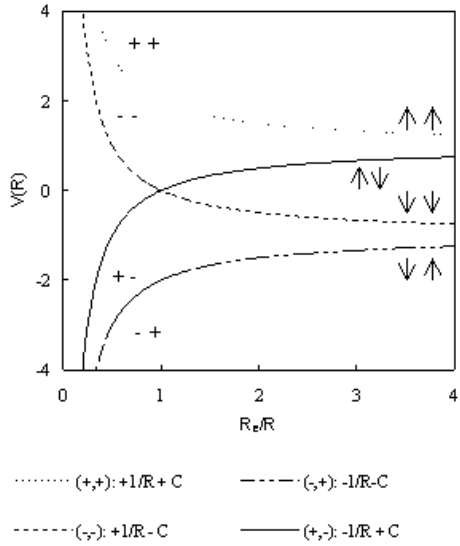

Fig. 5 Gauge symmetry and Coulomb's law

......... (+,+): +1/R + C      − − − − (−,+): −1/R − C
− − − − − (−,−): +1/R − C      ———— (+,−): −1/R + C

Fig. 2 (Fig. 3a from ref [18])

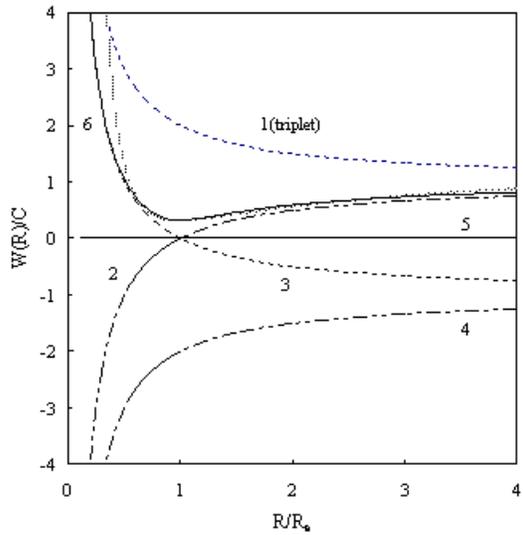

Fig. 3a Potential n = −1

1: +1+1/R; 2: +1−1/R; 3: −1+1/R; 4: −1−1/R; 5: Kratzer; 6: perturbed Coulomb

Fig. 3 (Fig. 10a from ref [18])

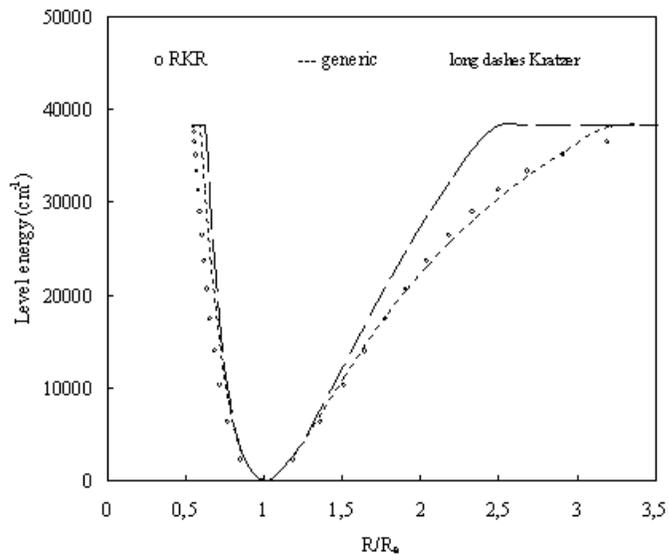

Fig. 4 (Fig. 10b from ref [18])

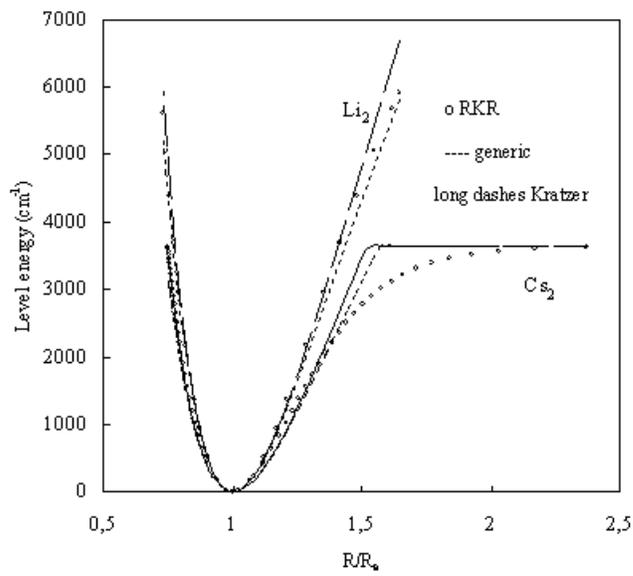

Fig. 5 (Fig. 10c from ref [18])

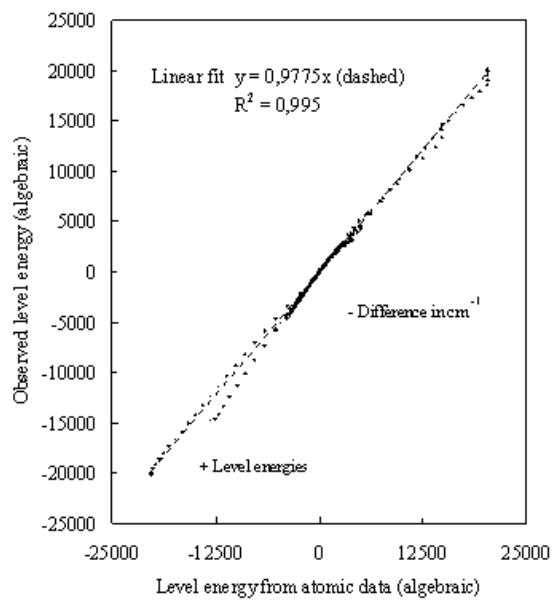

Fig. 10c Observed versus level energies computed from atomic data for 8 bonds (300 data points)